\documentclass[prl,twocolumn,showpacs,groupedaddress]{revtex4}

\def\ba{\begin{eqnarray}}
\def\ea{\end{eqnarray}}
\def\be{\begin{equation}}
\def\ee{\end{equation}}

\usepackage{epsfig,color}
\usepackage{dcolumn}

\begin{document}

\title{Magnetism, superconductivity, and pairing symmetry in Fe-based superconductors}
\author {A.V. Chubukov$^{1}$, D.V. Efremov$^{2}$, and I. Eremin$^{3,4}$}
\affiliation {$^1$
 Department of Physics,
 University of Wisconsin-Madison, Madison, WI 53706, USA \\
$^2$ Institut f\"ur Theoretische Physik, Technische
Universit\"at Dresden, 01062 Dresden, Germany\\
$^3$ Max-Planck Institut f\"ur Physik komplexer
Systeme, D-01187 Dresden, Germany \\
$^4$ Institute f\"ur Mathematische und Theoretische  Physik,
TU-Braunschweig, D-38106 Braunschweig, Germany}
\date{\today}

\begin{abstract}
We analyze antiferromagnetism and superconductivity in novel
$Fe-$based superconductors within the itinerant model of small
electron and hole pockets  near $(0,0)$ and $(\pi,\pi)$. We argue
that the effective interactions in both channels logarithmically
flow towards the same values at low energies, {\it i.e.},
antiferromagnetism and superconductivity must be treated on equal
footings. The magnetic instability comes first for equal sizes of
the two pockets, but looses to superconductivity upon doping. The
superconducting gap has no nodes, but changes sign between the two
Fermi surfaces (extended $s$-wave symmetry). We argue that the $T$
dependencies of the spin susceptibility and NMR relaxation rate for
such state are exponential only at very low $T$, and can be well
fitted by power-laws over a wide $T$ range below $T_c$.
\end{abstract}


\maketitle

\section{Introduction}

Recent discovery of superconductivity in the iron-based layered
pnictides  with $T_c$ ranging between $26$ and $52$K  generated
enormous interest  in the physics of these
materials~\cite{kamihara,chen1,chen2,ren,rotter}. The
superconductivity has been discovered in oxygen containing RFeAsO
(R=La, Nd, Sm) as well as in oxygen free AFe$_2$As$_2$ (A=Ba, Sr,
Ca). Like the cuprates, the pnictides are  highly two-dimensional,
their parent material shows antiferromagnetic long-range order below
150K \cite{kamihara,dong,cruz,nomura,klauss}, and superconductivity
occurs upon doping of either electrons
\cite{kamihara,chen1,chen2,ren} or holes \cite{rotter} into the FeAs
layers.

The close proximity of antiferromagnetism and superconductivity
fueled early speculations that the physics of the pnictides is
similar to the cuprates, and involves insulating
behavior\cite{Si,Daghofer,Ma1}. However, there is a growing
consensus among researchers that Mott physics does not play a
significant role for the iron pnictides, which remain itinerant for
all doping levels, including parent compounds, in which magnetic
order is of spin-density-wave (SDW) type rather than Heisenberg
antiferromagnetism of localized spins\cite{Raghu,Hirschfeld1}. This
is evidenced by, e.g.,  a relatively small value of the observed
magnetic moment per Fe atom, which is around $12-16\%$ of $2\mu_B$
\cite{klauss,cruz}. In another distinction to the cuprates,
electronic structure proposed by band structure
calculations\cite{Lebegue,Singh,Boeri,Mazin,Kuroki} and supported by
ARPES~\cite{kaminski,feng} consists of two small hole pockets
centered around  $\Gamma$ point (${\bf p} = (0,0)$) and two small
electron pockets centered around $M$ point (${\bf p} = {\bf Q}
=(\pi,\pi)$) in the {\it folded} Brillouin zone (BZ) (two $Fe$
artoms in the unit cell, we set interatomic spacing $a=1$)
\begin{figure}[tbp]
\includegraphics[angle=0,width=0.7\linewidth]{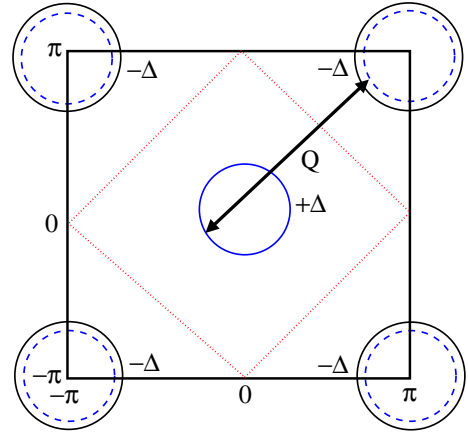}
\caption{(color online) A simplified FS geometry of doped Fe-based
superconductors, used in the present work. At zero doping, the Fermi
surface consists of an electron pocket  around $(\pi, \pi)$ (black
solid curve), and a hole pocket of roughly equal size around $(0,0)$
(blue solid curve). We neglect in this work the fact that there are
two hole and two electron pockets. In this system, there is a
near-perfect nesting between hole and electron pockets (moving a
hole FS by $(\pi,\pi)$ one obtains a near-perfect match with an
electron FS). Upon electron doping, the size of the electron pocket
increases (dashed blue $\rightarrow$ black), and this breaks the
nesting. $+\Delta$ and $-\Delta$ are the values of the
superconducting gaps on the two FS for $s^+$ superconducting state.}
\label{fig1}
\end{figure}

In this paper, we address three issues for the pnictides: (i) what
interactions cause SDW order and superconductivity, (ii) what is the
gap symmetry, and (iii) what are the implications of the gap
symmetry for the experiments in the superconducting (SC) state.  We
argue that both magnetic and pairing instabilities are determined by
the same interband pair hopping which transforms two fermions near
the hole Fermi surface into two fermions near the electron Fermi
surface (and vice versa). This interaction may be weak at high
energies (of order bandwidth), but it flows under renorm-group (RG)
and ultimately determines the couplings in both SDW and Cooper
channels at low-energies. When electron and hole pockets are nearly
identical, SDW instability occurs at a higher $T$. When the
near-identity is broken by either hole or electron dopings, the
Cooper instability comes first. This pairing interaction sets the
gaps in hole and electron pockets to be of equal magnitude $\Delta$,
but of opposite signs (an extended $s-$wave symmetry, $s^+$). The
ratio $2\Delta/T_c = 3.53$ is, however, the same as in BCS theory,
as there is no angular variations of the gap along the FS.

A fingerprint of $s^+$ gap symmetry and near-equal electron and hole
pockets is the existence of a magnetic  collective mode inside the
gap for momenta near ${\bf Q}$ (a spin resonance), whose dispersion
$(\Omega^2_0 + (v^2_F/2) ({\bf q-Q})^2)^{1/2}$, where $v_F$ is the
Fermi velocity, closely resembles Anderson-Bogolubov mode in
uncharged superconductors. Another fingerprint is a strong reduction
of the nuclear magnetic resonance (NMR) relaxation rate $1/T_1$ in
the clean limit, due to vanishing of the coherence factor for
$\chi^{''} ({\bf q}, \omega)/\omega$ for ${\bf q = Q}$.  We argue
that in this situation, $1/T_1$ is predominantly due to impurities,
which are partly pair-breaking even when non-magnetic. We show that,
in the presence of impurity scattering, $1/T_1$ is exponential in
$T$ only for very low temperatures, and over a wide range of $T<T_c$
is well described by $1/T_1 \propto T^3$, as if the gap had nodes.
Over the same range of $T$, the uniform susceptibility is
near-linear in $T$.

Our results partly agree and partly disagree with some earlier works
on $Fe$-pnictides.  Mazin et al.~\cite{Mazin} and Gorkov and
Barzykin~\cite{gorkov_barz} conjectured that the pairing symmetry
should be $s^+$. Our results agree with theirs and also with Eremin
and Korshunov~\cite{eremin}, who analyzed numerically the magnetic
response at ${\bf Q}$ within RPA for an $s^+$ superconductor and
found the resonance peak below $2\Delta$. Cvetkovic and
Tesanovic~\cite{cvetkovic} noticed that for identical electron and
hole pockets, Cooper and particle-hole channels become
undistinguishable and should be treated equally -- the notion we
share. Wang et al~\cite{wang,wang_2} performed numerical RG study of the
pairing symmetry and found an $s^+$ gap symmetry for two-band model
and a conventional $s-$wave symmetry for five-band model. We can
only compare the results for the two-band model, for which we also
found an attraction in in $s^+$ channel. There is, however, an
important difference between our results and those of Wang et al. In
our case, the bare interaction in $s^+$ channel is repulsive, and
attraction emerges only below some energy scale, due to RG flow of
the coupling. In their analysis, the bare interaction is zero, and
attraction emerges already after an infinitesimal RG transformation.
Lorenziana et al. \cite{lorenziana} used unrestricted Hartree-Fock
approximation and studied possible phases that may compete with
superconductivity in FeAs layers. We found that SDW is the main
competitor, but CDW with complex order parameter is close second.

On experimental side, ARPES\cite{kondo,ding,kaminski_gap} and
Andreev spectroscopy\cite{chen} measurements have been interpreted
as evidence for a nodeless gap, while NMR data were argued to follow
$1/T_1 \propto T^3$ and were interpreted as evidence for a  $d-$wave
gap\cite{nakai,grafe} or multiple gaps\cite{matano}. Our results
show that the $T$ dependence of $1/T_1$ in a dirty $s^+$
superconductor mimics $T^3$ over a wide range of $T$ and become
exponential only at very low temperatures.

\section{The model}

We model iron pnictides by an itinerant electron system with two
electronic orbitals,  and we assume that the hybridization between
the orbitals leads to small  hole and electron pockets located near
$(0,0)$ and $(\pi,\pi)$, respectively in the folded BZ (two $Fe$
atoms per unit cell) (Fig.\ref{fig1}).  The extension to a more
realistic case of four (or even five) orbitals and two hole and two
electron pockets is straightforward, and does not lead to new
physics except for a magnetically ordered state, where four-pocket
structure is essential for proper identification of relative
magnetic ordering of spins of the two $Fe$ atoms from the unit cell
in folded BZ~\cite{subir_1,wang_2,com_rev}.

 We assume that electron-electron interaction is short-range
(Hubbard-like) and involves two couplings -- between fermionic
densities from the same orbital and from different
orbitals~\cite{phillips}.
 The Hamiltonian has the form ${\cal H} = {\cal H}_2 + {\cal H}_4$, where
\begin{widetext}
\begin{eqnarray}
&&{\cal H}_2 =  \sum_{{\bf p},\sigma} \epsilon_{1,\bf p}
\psi^\dagger_{1,{\bf p},\sigma} \psi_{1,{\bf p},\sigma} +
\epsilon_{2,\bf p} \psi^\dagger_{2,{\bf p},\sigma} \psi_{2,{\bf
p},\sigma} + \Gamma_{\bf p} \left(\psi^\dagger_{1,{\bf p},\sigma}
\psi_{2,{\bf p},\sigma} + \psi^\dagger_{2,{\bf
p},\sigma} \psi_{1,{\bf p},\sigma}\right) \nonumber \\
&& {\cal H}_4 =  \frac{U_{11}}{2} \sum_{{\bf p}_i,\sigma \neq \sigma'}\left[\psi^\dagger_{1,{\bf p}_1,\sigma} \psi_{1,{\bf p}_2,\sigma} \psi^\dagger_{1,{\bf p}_3,\sigma'}
\psi_{1,{\bf p}_4,\sigma'} +  \psi^\dagger_{2,{\bf p}_1,\sigma} \psi_{2,{\bf p}_2,\sigma} \psi^\dagger_{2,{\bf p}_3,\sigma'}  \psi_{2,{\bf p}_4,\sigma'}\right] +
U_{12}  \sum_{{\bf p}_i,\sigma,\sigma'}
\psi^\dagger_{1,{\bf p}_1,\sigma} \psi_{2,{\bf p}_2,\sigma}
\psi^\dagger_{2,{\bf p}_3,\sigma'}  \psi_{2,{\bf p}_4,\sigma'}, \nonumber\\
&&
 \label{eq:1_1}
\end{eqnarray}
\end{widetext}
where where ${\bf p_1} +{\bf p_2} = {\bf p_3} + {\bf p_4}$, and
$U_{11}$ is intra-orbital, and $U_{12}$ inter-orbital  interactions
which we approximate by momentum-independent (on-site) values.

The quadratic form can be easily diagonalized by
\begin{eqnarray}
\psi_{1,{\bf p},\sigma} &=& \cos{\theta_{\bf p}} c_{{\bf p},\sigma}
+ \sin{\theta_{\bf p}}
f_{{\bf p},\sigma} \nonumber\\
\psi_{2,{\bf p},\sigma} &=& \cos{\theta_{\bf p}} f_{{\bf p},\sigma} -
\sin{\theta_{\bf p}}
c_{{\bf p},\sigma}
\label{m_1}
\end{eqnarray}
with $\tan {2\theta_{\bf p}} = 2 \Gamma_{\bf p}/(\epsilon_{2,{\bf p}} - \epsilon_{1,{\bf p}})$. This yields
\begin{equation}
{\cal H}_2 = \sum_{{\bf p},\sigma} \epsilon^c_{\bf p} c^\dagger_{{\bf
p},\sigma} c_{{\bf p},\sigma} +  \epsilon^f_{\bf p} f^\dagger_{\bf
p,\sigma} f_{{\bf p},\sigma}, \label{eq:1}
\end{equation}
where
\begin{equation}
\epsilon^{c,f}_{\bf p} = \frac{\epsilon_{1,{\bf p}} + \epsilon_{2,{\bf p}}}{2} \mp \frac{1}{2} \sqrt{(\epsilon_{1,{\bf p}} - \epsilon_{2,{\bf p}})^2 + 4 \Gamma^2_{\bf p}}
\label{s_1}
\end{equation}
Under some conditions on the original dispersions $\epsilon_{1,{\bf
p}}$ and $\epsilon_{2,{\bf p}}$, and on the hybridization term
$\Gamma_{\bf p}$, the two bands of fermionic excitations form small
hole and electron pockets near $(0,0)$ and ${\bf Q} = (\pi,\pi)$,
with roughly equal size, as in the iron pnictides. This happens if,
e.g.,  $\epsilon_{1,{\bf p}}$ and $\epsilon_{2,{\bf p}}$ change sign
under ${\bf p}\rightarrow {\bf p} + {\bf Q}$,  $\epsilon_{1,{\bf 0}}
+ \epsilon_{2,{\bf 0}} >0$, $\Gamma_{{\bf p} + {\bf Q}} = \pm
\Gamma_{\bf p}$, and $ \sqrt{(\epsilon_{1,{\bf p}} -
\epsilon_{2,{\bf p}})^2 + 4 \Gamma^2_{\bf p}} \leq |\epsilon_{1,{\bf
p}} + \epsilon_{2,{\bf p}}|$. Under these conditions,
$\epsilon^c_{\bf p} = - \epsilon^f_{{\bf p} + {\bf Q}}$, and
$\epsilon^{c}_{\bf p}$ describes a hole band with the maximum of
energy at $(0,0)$, while $\epsilon^{f}_{\bf p}$ describes an
equivalent electron band with the minimum of energy at ${\bf Q}$.
Upon doping, chemical potential shifts, one Fermi surface gets
larger while the other gets smaller,  see Fig.\ref{fig1}.

\begin{figure}[tbp]
\includegraphics[angle=0,width=1\linewidth]{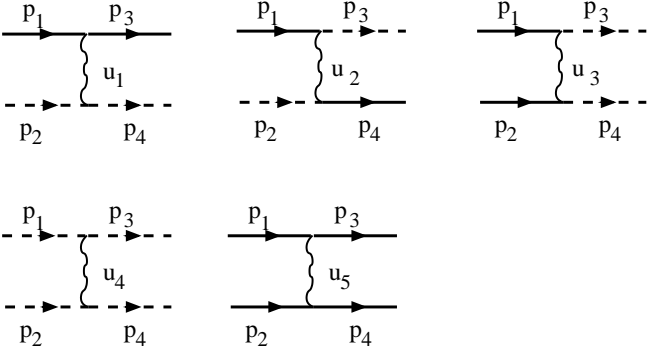}
\caption{Five relevant bare vertices. Solid and dashed lines
represent fermions from $c-$ band (near ${\bf k}=0$) and $f-$band
(near ${\bf k} = (\pi,\pi)$).} \label{fig2}
\end{figure}

In itinerant systems, the interactions are expected to be small
compared to the fermionic bandwidth, and the physics is dominated by
fermions near the Fermi surface (FS). The projection of the
Hubbard interaction term  ${\cal H}_{4}$ onto
$c$ and $f$ fermions leads to five different interactions:
\begin{widetext}
\begin{eqnarray}
&&H_{4}=U_1^{(0)} \sum  c^{\dagger}_{{\bf p}_3 \sigma}
f^{\dagger}_{{\bf p}_4 \sigma'}  f_{{\bf p}_2 \sigma'} c_{{\bf p}_1
\sigma}  +U_2^{(0)} \sum f^{\dagger}_{{\bf p}_3 \sigma}
c^{\dagger}_{{\bf p}_4 \sigma'} f_{{\bf p}_2 \sigma'} c_{{\bf p}_1
\sigma}  \nonumber \\
&& +
 \frac{U_3^{(0)}}{2}~ \sum
\left[f^{\dagger}_{{\bf p}_3 \sigma} f^{\dagger}_{{\bf p}_4 \sigma'}
c_{{\bf p}_2 \sigma'} c_{{\bf p}_1 \sigma} + h.c \right]
 ~+~ \frac{U_4^{(0)}}{2} \sum  f^{\dagger}_{{\bf p}_3 \sigma}
f^{\dagger}_{{\bf p}_4 \sigma'} f_{{\bf p}_2 \sigma'} f_{{\bf p}_1
\sigma} ~+~ \frac{U_5^{(0)}}{2}  \sum  c^{\dagger}_{{\bf p}_3
\sigma} c^{\dagger}_{{\bf p}_4 \sigma'} c_{{\bf p}_2 \sigma'}
c_{{\bf p}_1 \sigma} \label{eq:2}
\end{eqnarray}
\end{widetext}
where  the momenta of $c-$ fermions are near $(0,0)$, the momenta of
$f-$fermions are near $(\pi,\pi)$, and the momentum conservation is
assumed. We present these interactions graphically in
Fig.\ref{fig2}.

We label the couplings with subindex $''0''$ to emphasize that these
are the bare couplings.  The terms with  $U_4^{(0)}$ and $U_5^{(0)}$
are intraband interactions, the terms with $U^{(0)}_1$ and
$U^{(0)}_2$ are interband interactions with momentum transfer $0$
and ${\bf Q}$, respectively, and the term with  $U_3^{(0)}$ is
interband pair hopping.

Note that in our Fermi-liquid description, all vertices  in Eq.
(\ref{eq:2}) are $\delta$-functions in spin indices, {\it i.e.}, all
interactions are in the charge channel~\cite{landau}, and there are
no direct spin-spin interaction terms with spin matrices in the
vertices. However, if the original Hubbard interaction is on-site,
one can use another, equivalent, description in which Pauli
principle is build into the Hamiltonian, and the intra-orbital terms
with equal spin projections are  eliminated from the Hamiltonian. In
this description, $U^{(0)}_{1}$, $U^{(0)}_4$, and $U^{(0)}_5$ terms
appear as effective Hubbard interactions, while $U^{(0)}_2$ and
$U^{(0)}_3$ appear as a  magnetic,  Hund term~\cite{wang}.

In explicit form, $U^{(0)}_i$ are
\begin{eqnarray}
&& U^{(0)}_1 = \frac{1}{2} \left[(U_{11} + U_{12}) - \cos{2 \theta_0} \cos{2\theta_{\bf Q}} (U_{11} - U_{12})\right]\nonumber\\
&& U^{(0)}_{2,3} = \frac{U_{11}}{2} (1 -  \cos{2 \theta_0} \cos{2\theta_{\bf Q}}) - \frac{U_{12}}{2}  \sin{2 \theta_0} \sin{2\theta_{\bf Q}}\nonumber \\
&& U^{(0)}_4 = \frac{U_{11}+U_{12}}{2 } + \frac{U_{11} - U_{12}}{2} \cos^2{2 \theta_0} , \nonumber\\
&& U^{(0)}_5 = \frac{U_{11}+U_{12}}{2 } + \frac{U_{11} - U_{12}}{2} \cos^2{2 \theta_{\bf Q}} ,~ \label{s_2}
\end{eqnarray}

For the case that we considered above ($\epsilon_{1,{\bf
p}} = - \epsilon_{1,{\bf p} + {\bf Q}}, \epsilon_{2,{\bf
p}} = - \epsilon_{2,{\bf p} + {\bf Q}},
\Gamma_{{\bf p} + {\bf Q}} = \pm \Gamma_{\bf p}$), we have $\theta_{{\bf Q}}
= \pi/2 \mp \theta_0$,  we have
\begin{eqnarray}
&& U^{(0)}_1 = U^{(0)}_4 = U^{(0)}_5 = \frac{U_{11}+U_{12}}{2 } + \frac{U_{11} - U_{12}}{2} \cos^2{2 \theta_0} , \nonumber\\
&& U^{(0)}_2 = U^{(0)}_3 =  \frac{U_{11}}{2} (1 + \cos^2{2 \theta_0}) \mp
 \frac{U_{12}}{2}  \sin^2{2 \theta_0}
\label{ya_1}
\end{eqnarray}
We assume that the intra and inter-orbital Hubbard-type interactions
$U_{11}$ and $U_{12}$ are positive (repulsive). We see from
(\ref{ya_1}) that density-density couplings $U^{(0)}_1, U^{(0)}_4$,
and $U^{(0)}_5$ are  positive and the largest. The couplings
$U^{(0)}_2$ and $U^{(0)}_3$ are smaller for the case when the
hybridization term is even under ${\bf p} \rightarrow {\bf p} + {\bf
Q}$, i.e., $\Gamma_{\bf p} = \Gamma_{{\bf p} + {\bf Q}}$, and are
the same as $U^{(0)}_1, U^{(0)}_4$, and $U^{(0)}_5$ when
$\Gamma_{\bf p} = \Gamma_{{\bf p} + {\bf Q}}$. The first case
corresponds to on-site hybridization and is more realistic that the
second one, which requires hybridization to involve predominantly
nearest neighbors. Below we will consider only the first case
$\Gamma_{\bf p} = \Gamma_{{\bf p} + {\bf Q}}$. Note that in this
situation,  the sign of $ U^{(0)}_2 = U^{(0)}_3$ depends on
$\theta_0$ and on the relative strength of the intra-orbital and
inter-orbital Hubbard terms. If $U_{11} > U_{12} \sin^2{2
\theta_0}/(1 + cos^2{2\theta_0})$, these couplings are positive, if
$U_{11} < U_{12} \sin^2{2 \theta_0}/(1 + cos^2{2\theta_0})$, they
are negative. A more likely situation is when the intra-orbital
Hubbard term $U_{11}$ is larger than inter-orbital $U_{12}$, in
which case $U^{(0)}_2$ and $U^{(0)}_3$ are positive.

For convenience, below we will be using dimensionless interactions
\begin{equation}
u_i = U_i N_0
\label{m_3}
\end{equation}
where $N_0$ is the fermionic density of states (DOS) which we approximate
by a constant. For itinerant systems, $|u^0_i| <1$ and can be
treated within Fermi liquid theory. We will also count the momenta
of $f-$fermions as deviations from ${\bf Q}$ ($f_{\bf p} \rightarrow
f_{{\bf p} +{\bf Q}}$) in which case all running momenta in the
vertices are small.

\subsection{Density wave and pairing instabilities}

We searched for  possible density-wave and Cooper-pairing
instabilities for our model, and found that the ones which may
potentially occur are spin density wave (SDW) and charge density
wave (CDW) instabilities with momentum ${\bf Q}$ and with either
real or imaginary order parameter, and superconducting (SC)
instability either in pure $s$ channel (the gaps $\Delta_c$ and
$\Delta_f$ have the same sign), or in $s^+$ channel (the gaps
$\Delta_c$ and $\Delta_f$ have opposite sign). Density-wave
instabilities with ${\bf q}=0$ and pairing instabilities with ${\bf
q=Q}$ do not occur within our model because the corresponding
kernels vanish for a constant DOS. The instabilities with
momentum-dependent order parameter, like a nematic
instability~\cite{subir_1} also do not occur simply because we set
all interactions to be momentum-independent and weak, and will
neglect regular (non-logarithmic) corrections which give rise to
 the momentum dependence of the scattering amplitude in a Fermi
liquid~\cite{landau}.

\begin{figure}[tbp]
\includegraphics[angle=0,width=1\linewidth]{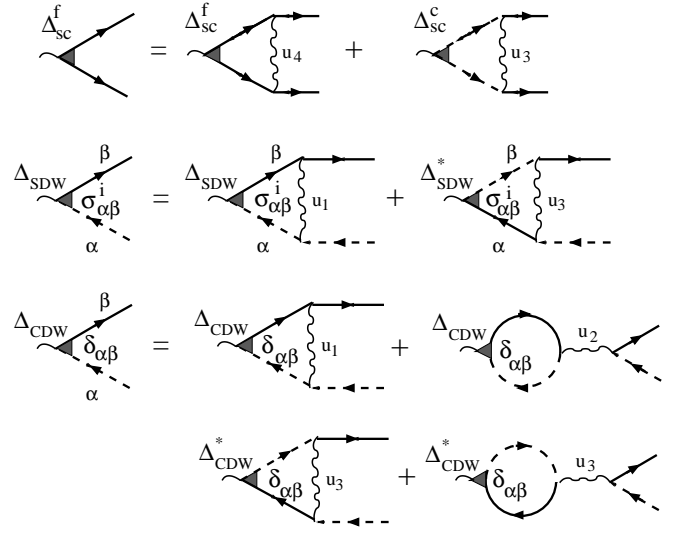}
\caption{Diagrammatic representation of the equations for SDW, CDW,
and SC instability temperatures. The equations for density-wave
$T^{(r,i)}_{sdw}$ and  $T^{(r,i)}_{cdw}$ are obtained by adding and
and subtracting equations for $\Delta_{sdw}$ and $\Delta^*_{sdw}$,
and for $\Delta_{cdw}$ and $\Delta^*_{cdw}$, respectively. The
equations for $T^{(s)}_{sc}$ and  $T^{(s^+)}_{sc}$  obtained by
adding and and subtracting equations for $\Delta^c$ and $\Delta^f$.}
\label{fig3}
\end{figure}

The temperatures of potential density-wave and pairing instabilities
are obtained by conventional means, by introducing infinitesimal
couplings
\begin{eqnarray}
&& \Delta_{sdw} \sum_k c^\dagger_{{\bf k},\alpha}
\sigma^z_{\alpha\beta} f_{{\bf k+Q},\beta}, \nonumber \\
&&  \Delta_{cdw} \sum_k c^\dagger_{{\bf k},\alpha}
\delta_{\alpha\beta} f_{{\bf k+Q},\beta}, \nonumber \\
&&\Delta^c_{sc} \sum_k c_{{\bf
k},\alpha}\sigma^y_{\alpha\beta}c_{-{\bf k},\beta} + \Delta^f_{sc}
\sum_k f_{{\bf k+Q},\alpha}\sigma^y_{\alpha\beta}f_{{\bf
-k-Q},\beta} \nonumber\\
\end{eqnarray}
with complex $\Delta_{sdw},~\Delta_{cdw}$, and real
$\Delta^{c,f}_{sc}$, and analyzing when the response functions
diverge.  We label the corresponding instability temperatures as
$T^{(r,i)}_{sdw}$ $T^{(r,i)}_{cdw}$ and $T^{(s,s^+)}_{sc}$, where
$r,i$ mean real or imaginary density-wave order parameter, and $s,
s^+$ mean $s-$wave or extended $s-$wave, respectively. The
linearized equations for the order parameters are presented
graphically in Fig.\ref{fig3}.  They have non-zero solutions when
\begin{eqnarray}
1 &=& -T^{(r,i)}_{sdw} \sum_{\omega_m} \Gamma^{(r,i)}_{sdw} \int d \epsilon_{\bf k} G^c_{{\bf k}\omega_m} G^f_{{\bf k+Q},\omega_m} \nonumber \\
1 &=& -T^{(r,i)}_{cdw} \sum_{\omega_m} \Gamma^{(r,i)}_{cdw} \int d \epsilon_{\bf k} G^c_{{\bf k}\omega_m} G^f_{{\bf k+Q},\omega_m} \nonumber \\
1 &=& -T^{(s,s^+)}_{sc} \sum_{\omega_m} \Gamma^{(s,s^+)}_{sc} \int d
\epsilon_{\bf k} G^c_{{\bf k}\omega_m} G^c_{-{\bf k},-\omega_m}~~
\label{a_1}
\end{eqnarray}
Here
\begin{eqnarray}
\Gamma^{(r,i)}_{sdw} &=& u_1 \pm u_3, ~~\Gamma^{(r,i)}_{cdw} =
u_1 \mp u_3 -2 u_2,\nonumber \\
\Gamma^{(s)}_{sc} &=& u_4 + u_3, ~~\Gamma^{(s^+)}_{sc} = u_4 - u_3
\label{m_4}
\end{eqnarray}
are the full interactions  in the SDW, CDW,  and  SC channels. Eq.
(\ref{a_1}) is only valid for the largest instability temperature.
Below such $T$, the ordering in one channel affects susceptibilities
in the other channels.

For the bare parameters as in (\ref{ya_1})
\begin{eqnarray}
&&\Gamma^{(r)}_{sdw} = u^0_1 + u^0_3 \approx u_{11} (1 + \cos^2{2\theta_0}), \nonumber \\
&&\Gamma^{(i)}_{sdw} = u^0_1 - u^0_3 \approx u_{12} \sin^2 {2\theta_0} \nonumber \\
&& \Gamma^{(r)}_{cdw} = u^0_1 - u^0_3 -2 u^0_2 \approx 2u_{12} \sin^2{2\theta_0} - u_{11} (1 + \cos^2{2\theta_0}),\nonumber \\
&& \Gamma^{(i)}_{cdw} = u^0_1 + u^0_3 -2 u^0_2 \approx u_{12} \sin^2{2\theta_0}\nonumber \\
&& \Gamma^{(s)}_{sc}  = u^0_4 +  u^0_3 \approx u_{11} (1 + \cos^2{2\theta_0}), \nonumber \\
&& \Gamma^{(s^+)}_{sc}  = u^0_4 -  u^0_3 \approx u_{12} \sin^2{2\theta_0}
\label{tu_1}
\end{eqnarray}
where $u_{11} = U_{11} N_0$, $u_{12} = U_{12} N_0$ are dimensionless
intra-orbital and inter-orbital Hubbard couplings. The Stoner-like
SDW and CDW instabilities require $\Gamma_{sdw}, ~\Gamma_{cdw} >0$.
 At the bare level, $\Gamma^{(r)}_{sdw}$ is the largest positive
interaction when $u_{11} (1 + \cos^2{2\theta_0}) > u_{12}
\sin^2{2\theta_0}$ and $\Gamma^{(r)}_{cdw}$ is the largest positive
interaction when $u_{11}  (1 + \cos^2{2\theta_0}) < u_{12}
\sin^2{2\theta_0}$ {\it i.e.}, the system undergoes a conventional
SDW or CDW instability.   The SC instabilities requires an
attraction (a negative $\Gamma^{(s,s^+)}_{sc}$) and do not occur at
this level because both $\Gamma^{(s)}_{sc}$ and $
\Gamma^{(s^+)}_{sc}$ are {\it positive}.

\section{RG flow}

\begin{figure}[tbp]
\includegraphics[angle=0,width=1\linewidth]{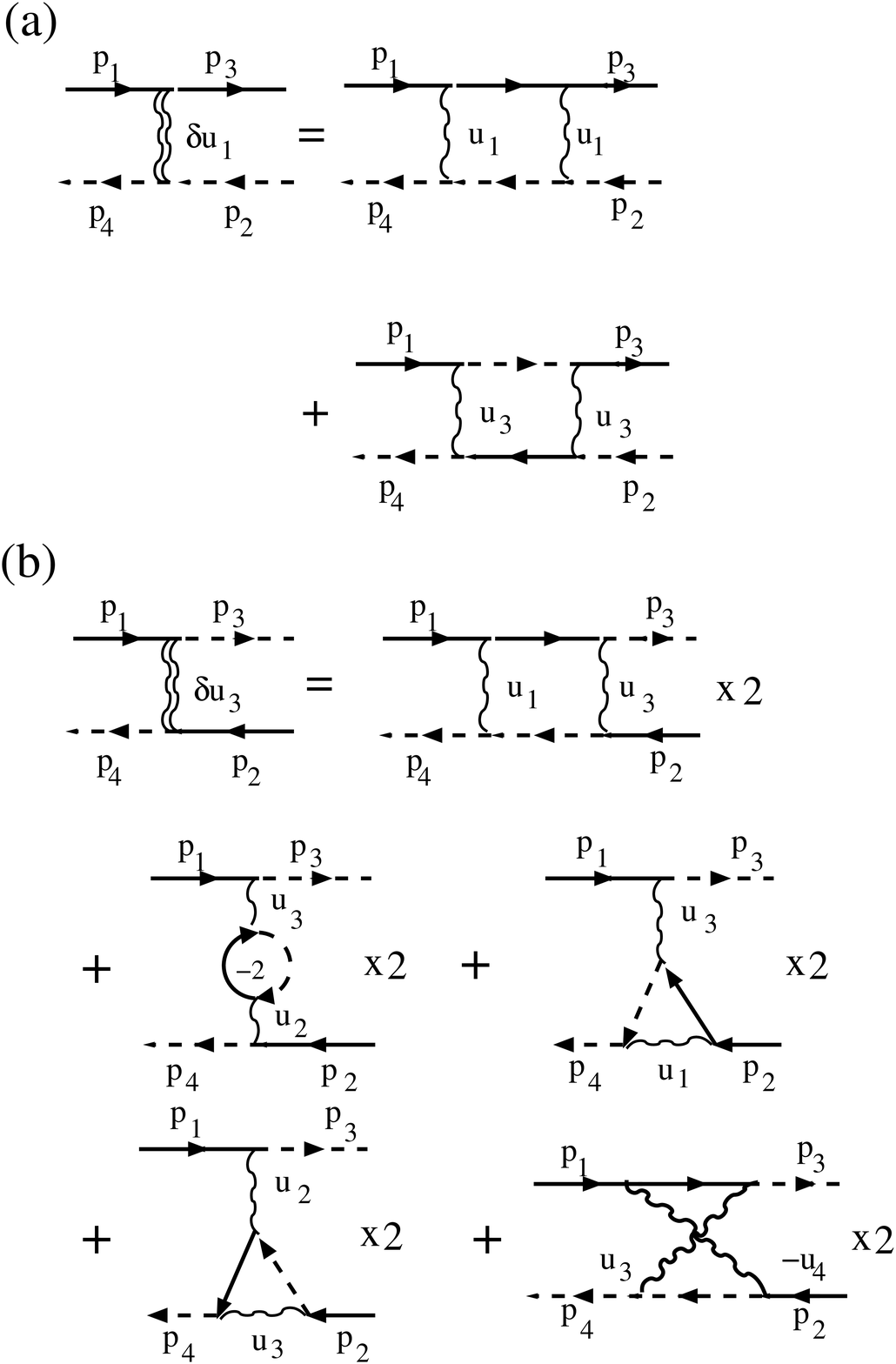}
\caption{Diagrams for one-loop vertex renormalizations. The
renormalizations of $u_1$ and $u_3$ are shown, others are obtained
in a similar way.}\label{fig2_1}
\end{figure}
Beyond mean-field, the potential SDW and SC instabilities  are
determined by $u_i$ at energies below the Fermi energy $E_F$, and
generally differ from  bare  $u^0_i$ defined  at energies comparable
to the bandwidth, $W$. For small size of the FS, $W >> E_F$, and the
intermediate range is quite large.  At $u^0_i <1$ the
renormalization can be considered in one-loop approximation. The
one-loop  diagrams, shown in Fig.\ref{fig2_1}, contain
particle-particle and particle-hole bubbles. The external momenta in
these diagrams are of order running $E \geq E_F$, while internal
momenta are generally of order $W$, {\it i.e.}, much larger. In this
situation, the dependence on the directions of the external momenta
is lost, {\it i.e.}, a SC vertex with zero total momentum and an SDW
vertex with transferred momentum ${\bf Q}$ are renormalized in the
same way. The crucial element of our analysis is the observation
that, for $\epsilon^c_{\bf p} = - \epsilon^f_{\bf p+q}$,
particle-hole channel is undistinguishable from particle-particle
channel, such that the renormalization in both channels are
logarithmical and interfere with each other. The presence of the
logarithms in both channels implies that the one-loop perturbation
theory must be extended to one loop RG analysis for the running
$u_i$ (in the diagrammatic language, one needs to sum up series of
logarithmically divergent parquet diagrams). The derivation of the
RG equations is straightforward (see Fig. \ref{fig2_1}). Collecting
combinatoric pre-factors for the diagrams, we obtain
\begin{eqnarray}
&&\dot{u}_1 = u_1^2 + u_3^2 \nonumber \\
&&\dot{u}_2 = 2 u_2(u_1 - u_2 ) \nonumber \\
&&\dot{u}_3 = 2 u_3(2u_1 - u_2-u_4 )\nonumber \\
&&\dot{u}_4 = -u_3^2 - u_4^2
\label{2}
\end{eqnarray}
where the derivatives are with respect to $\log W/E$, and $E$ is the
running energy scale. Similar, though not identical equations have
been obtained in the weak-coupling studies of the cuprates with the
``$t-$only'' dispersion~\cite{dzyal}.

\begin{figure}[tbp]
\includegraphics[angle=0,width=0.9\linewidth]{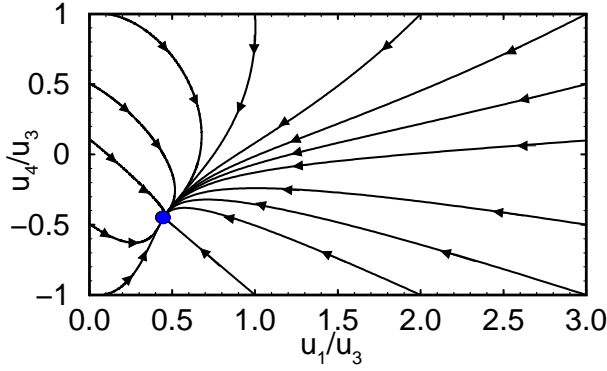}
\caption{(color online)
The RG flow  of Eqn. (\protect\ref{2}) in
variables $u_4/u_3$ and $u_1/u_3$.  The fixed point is $u_4/u_3 =
-1/\sqrt{5},~~u_1/u_3 = 1/\sqrt{5}$. The fourth variable, $u_2$,
becomes  small  near the fixed point compared to the other $u_i$,
We used  boundary conditions
$u^0_1=u^0_4$,  $u^0_3 = 0.1 u^0_1$, and for simplicity set $u^2_0 =0$.
}
\label{fig3_1}
\end{figure}
\begin{figure}[tbp]
\includegraphics[angle=0,width=0.9\linewidth]{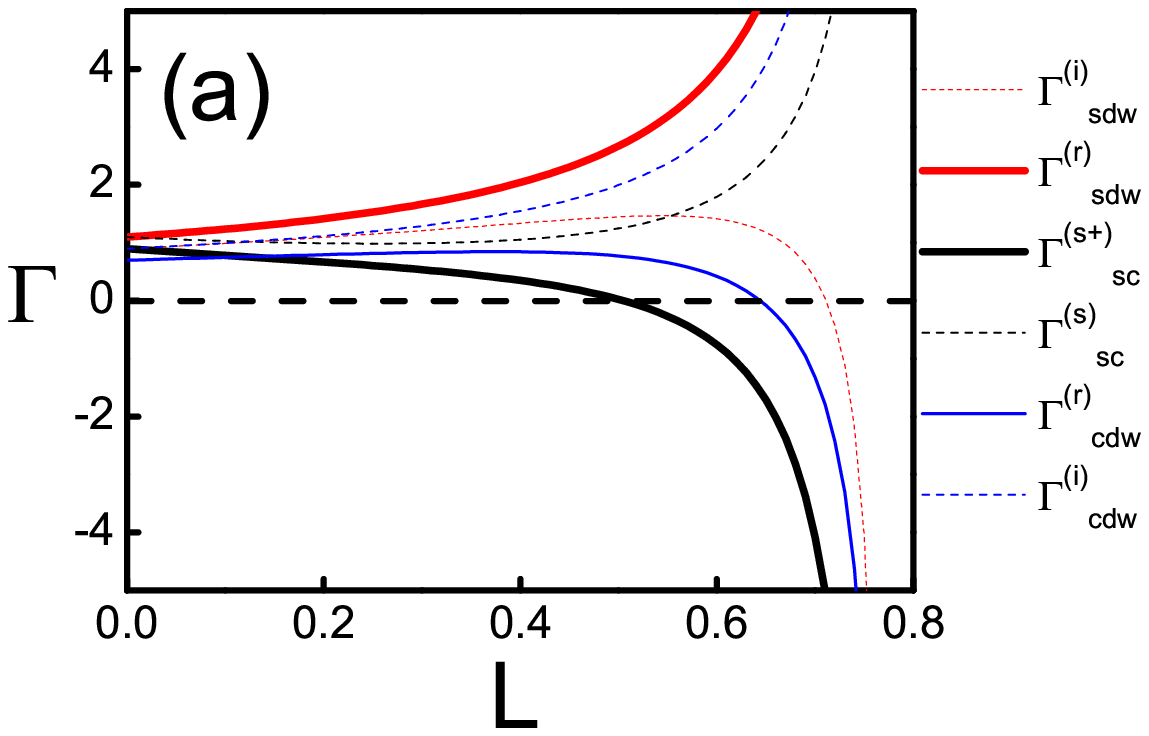}
\includegraphics[angle=0,width=0.9\linewidth]{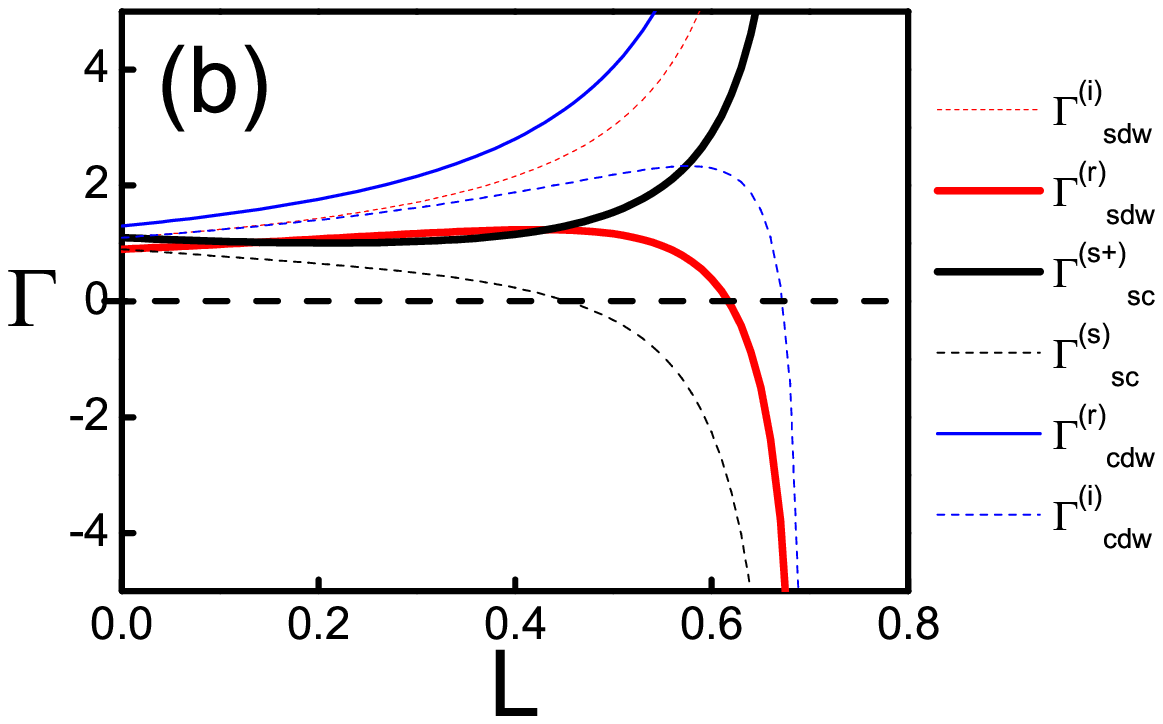}
\caption{(color online) The RG flow of the effective couplings in
various density-wave and superconducting channels vs $L = u^0 \log
W/E$. The boundary conditions are $u^0_1=u^0_4 = u_0$, and $u^0_2 =
u^0_3 = \pm 0.1 u_0$. The running couplings are in units of $u_0$.
Panel a) - the RG flow for $u^0_3 >0$. The extended $s-$wave ($s^+$)
superconducting channel becomes attractive above some $L$. The
strongest density-wave instability is in SDW channel, for real order
parameter.   Panel b) - the same for $u^0_3 <0$. The conventional
$s$-wave superconducting channel becomes attractive above some $L$.
The strongest density-wave instability is in CDW channel, again for
real order parameter.} \label{fig3_2}
\end{figure}

We see from Eq.(\ref{2}) that the pair hopping term $u_3$ is not
generated by other interactions, {\it i.e.}, $u_3 =0$ if $u^0_3 =0$.
In the absence of $u_3$, $\Gamma_{sdw}$ and $\Gamma^{s,s^+}_{sc}$
 -- $\Gamma_{sdw} = u_1$ increases and drives $T_{sdw}$ up, while
$\Gamma^{s,s^+}_{sc} = u_4$ logarithmically decreases, as it is
expected  for a  repulsive interaction~\cite{mcmillan}.
However, once $u^0_3$ is
finite, the system moves into the basin of attraction of another
fixed point, at which
\begin{equation}
u_3  \propto \frac{u}{1 - |u|
\ln{\frac{W}{E}}}, ~~u_1 = - u_4 =  \frac{|u_3|}{\sqrt{5}}, u_2 \propto |u_3|^{1/3}
\label{a_10}
\end{equation}
where $u$ depends on the bare values of the couplings. In Figs.\ref{fig3_1} and \ref{fig3_2} we
show the RG flow obtained by the numerical solution of Eq.
(\ref{2}).

The two key features of the new fixed point are (i) $|u_3|$ rapidly
increases and eventually becomes larger than $u_1$ by a factor
$\sqrt{5}$, and (ii) $u_4$ decreases, passes through zero, changes
sign, and then increases by magnitude and approaches
$-|u_3|/\sqrt{5}$ (see  Fig. \ref{fig3_1})

For positive $u^0_3 \approx u^0_2$, these results imply that
$\Gamma^{r}_{sdw} = u_1 + u_3$ remains positive and the largest out
of density-wave vertices {\it i.e.}, the highest density-wave
instability is a conventional SDW instability (see Fig.
\ref{fig3_2}a). Note, however, that $\Gamma^{(i)}_{cdw}$ is close
second as it only differs by  $u_2$ which under renormalization
becomes relatively small compared to $u_1$ and $u_3$ ($u_2 \propto
(u_3)^{1/3}$). The interaction in the $s^+$ SC channel,
$\Gamma^{(s+)}_{sc} = u_4-u_3$, becomes negative (attractive) below
some scale (Fig. \ref{fig3_2}a), while $\Gamma^{(s)}_{sc}$ remains
repulsive. We emphasize that the density-density vertex $u_4$
changes sign under renormalization, becomes attractive and also
supports SC. Moreover, the interactions in the SDW and the $s+$ SC
channel $\Gamma^{(r)}_{sdw} = u_1 + u_3$ and  $\Gamma^{(s+)}_{sc} =
u_3 - u_4$, become comparable to each other and eventually flow to
the same value $u_3 (1 + 1/\sqrt{5})$. The implication is that the
SDW order and $s^+$ superconductivity are {\it competing orders,
determined by effective interactions of comparable strength}.

For negative $u^0_3, u^0_2$, $\Gamma^{(r)}_{cdw} = u_1 + |u_3| + 2
|u_2|$ is the strongest, positive, density-wave vertex, and
$\Gamma^{(i)}_{sdw} = u_1 + |u_3|$ is a close second (see Fig.
\ref{fig3_2}b). $\Gamma^{(s^+)}_{sc} =  u_4 + |u_3|$ is now positive
(repulsive), but  $\Gamma^{(s)}_{sc} = u_4-|u_3|$, changes sign
under the renormalization and becomes negative (attractive), see
Fig. \ref{fig3_2}b. This implies that CDW now competes with a
conventional $s-$wave SC. Near the fixed point,  the interaction in
the $s-$channel $ \Gamma^{(s)}_{sc} \approx - |u_3| (1 +
1/\sqrt{5})$ is now larger than $\Gamma^{(r)}_{cdw} \approx |u_3| (1
-1/\sqrt{5})$ which implies that in this case $s$-wave SC likely
wins over CDW.

The generalization  of this analysis to 4-band model (or even five)
is straightforward and yields qualitatively similar behavior.

\subsection{Competing orders}

We next analyze in more detail Eqs. (\ref{a_1}) for $u^0_3 >0$. By
construction, the upper limit of the integration over internal
energies there is $O(E_F)$ as the contributions from higher energies
are already absorbed into the renormalized vertices. When hole and
electron Fermi surfaces are near-identical, i,e., $\epsilon^c_{\bf
k} = -\epsilon^f_{\bf k+Q}$ holds down to the lowest energies, both
SDW and SC susceptibilities are logarithmic in $T$
\begin{eqnarray}
&&- T \sum_\omega \int d \epsilon_{\bf k} G^c_{{\bf k}\omega_m}
G^f_{{\bf k+Q},\omega_m} =T \sum_\omega \int d \epsilon_{\bf k}
G^c_{{\bf k}\omega_m} G^c_{-{\bf k},-\omega_m} \nonumber \\&& =
\int_0^{E_F} \tanh\left(\frac{\omega}{2T}\right)
\frac{d\omega}{\omega} = \log \frac{E_F}{T} \label{a_6}
\end{eqnarray}
and from (\ref{a_1}) the largest instability temperature is either
\begin{equation}
T^{(r)}_{sdw} \sim E_F e^{-\frac{1}{\Gamma^{(r)}_{sdw}}},~~\mbox{or}
\quad T^{(s^+)}_{sc} \sim E_F
e^{-\frac{1}{\left|\Gamma^{(s^+)}_{sc}\right|}}.
\end{equation}
As $\Gamma^{(r)}_{sdw}$ is still larger than $\Gamma^{(s^+)}_{sc}$, the SDW
instability comes first. This is what, we believe, happens at zero
doping. Whether SC emerges as an extra order at a smaller $T$
requires a separate analysis as  the pairing susceptibility changes
in the presence of the SDW order. At a finite doping, all evidence
is that the two FS become unequal, {\it i.e.}, the condition
$\epsilon^c_{\bf k} = -\epsilon^f_{\bf k+Q}$ breaks down. In this
situation, the log $1/T$ behavior of the SDW polarization is cut,
and $T^{(r)}_{sdw}$ decreases and eventually becomes smaller than
$T^{(s^+)}_c$. At larger dopings, $T^{(s^+)}_c$ remains roughly doping
independent, while magnetic correlations decrease.

A remark about the SDW state. In the coordinate frame associated
with folded BZ,  $Fe$ ions are located at ${\bf r}_1 = (n_x, n_y)$,
where $n_x, n_y$ are integers (we recall that we set interatomic
spacing to one), but also at ${\bf r}_2 = (n_x +1/2, n_y + 1/2)$.
SDW instability with ${\bf Q} =(\pi,\pi)$ order
antiferromagnetically spins within the sublattice where ${\bf r} =
{\bf r}_1$, and within the sublattice where where ${\bf r} = {\bf
r}_2$, but do not fix relative orientation between the spins in the
two sublattices. To obtain full spin structure, we would need to
analyze spin ordering within full four-band structure (two electron
and and two hole orbitals), or go back into unfolded Brillouin zone.
For localized spins, this type of order is described by $J_1-J_2$
model for $J_2 > 0.5 J_1$. In the classical model, the angle between
${\bf r_1}$ and ${\bf r}_2$ sublattices is arbitrary, but quantum
fluctuations select $(0,\pi)$ or $(\pi, 0)$
state~\cite{quantum,coleman}. There is then an extra Ising degree of
freedom, which was argued~\cite{coleman,subir_1} to remain broken
even at $T > T_{sdw}$, when $SU(2)$ spin symmetry is restored.

\section{Superconducting state}

The SC $s^{+}$ state that we found has two features similar to a
conventional isotropic $s$-wave state. First, the superconducting
gaps on the hole and electron FS are opposite in sign, but equal in
magnitude. They, however, become unequal when $E_F$ on the two FS
become is different, which happens once the doping increases (or
when intraband density-density interactions $u_4$ and $u_5$ become
unequal). Second, solving the non-linear gap equation, we
immediately find that the gap $\Delta$ obeys the same BCS relation
$2\Delta = 3.53 T_c$ as for an isotropic $s-$wave state simply
because the pairing kernel contains either two $c-$fermions or two
$d-$fermions, but no $cf$ pairs.

The $s^+$ and $s$ SC states, however, differ qualitatively in the
presence of non-magnetic impurities. For $s-$state, non-magnetic
impurities do not affect $T_c$ and non-linear gap
equation\cite{hirschfeld11}. For $s^+$ state, the impurity potential
$U_{i}({\bf q})$ has intra and interband components $U_i(0)$ and
$U_i (\pi)$, respectively. The $U_i (\pi)$ components scatter
fermions with $+\Delta$ and $-\Delta$ and acts as a ``magnetic
impurity''\cite{golubov,preosti}. Specifically, for the $s^+$ state,
normal and anomalous Greens functions in the presence of impurities
are
\begin{eqnarray}
&&G^{c,f}_{{\bf k}, \omega_m} = \frac{Z_{\omega_m} \omega_m \pm
\epsilon_{\bf k}}{
Z^2 _{\omega_m}(\omega^2_m + {\bar \Delta}^2_{\omega_m}) + \epsilon^2_{\bf k}} \nonumber \\
&&  F^{c,f}_{{\bf k}, \omega_m} = \pm \frac{Z_{\omega_m} {\bar
\Delta_{\omega_m}}}{Z^2 _{\omega_m}(\omega^2_m + {\bar \Delta}^2_{\omega_m}) +
\epsilon^2_{\bf k}}, \label{a_9}
\end{eqnarray}
and the fermionic $Z = 1 + \Sigma (\omega_m)/\omega_m$ and the
renormalized gap ${\bar \Delta}_{\omega_m}$ in the
Born approximation are given by
\begin{eqnarray}
&&Z = 1 + \frac{U_{i} (0) + U_i (\pi)}{\sqrt{{\bar \Delta}^2 +\omega^2_m}} \nonumber \\
&&\frac{\bar \Delta_{\omega_m}}{\Delta} -1 = -
 \frac{b_T {\bar \Delta}_{\omega_m}}{\sqrt{{\bar \Delta}^2_{\omega_m} + \omega^2_m}},
\label{a_4}
\end{eqnarray}
where $\Delta = \Delta (T)$ is the frequency-independent
order parameter, and $b_T
= 2U_i(\pi)/\Delta (T)$. Below we use $b_{T=0} = b$ as a measure of
the strength of impurity scattering. Note that $b$ is a complex function
 of the impurity strength as the order parameter is also affected by
 impurities (see below).

For $U_i(\pi) =0$, ${\bar \Delta} = \Delta$, i.e., superconductivity
is not influenced by impurities. For $U_i (\pi) \neq 0$, ${\bar
\Delta}_{\omega_m}$ becomes frequency dependent, as if the
impurities were magnetic. At $T=0$, and $b \geq 1$, the system
displays gapless superconductivity\cite{abrgorkov}:  in real
frequencies ${\bar \Delta}_{\omega_m} \propto -i \omega$ at small
$\omega$, and the DOS at zero energy acquires a finite value
$N(\omega =0) = (1 - (1/b)^2)^{1/2}$. Superconductivity at $T=0$
eventually disappears when $\Delta$ vanishes, {\it i.e.}, when $b$
tends to infinity.

\begin{figure}[t]
\includegraphics[angle=0,width=1.0\linewidth]{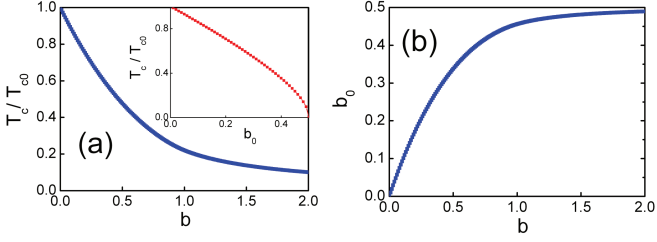}
\caption{ T$_c$ (a) and  $b_0 = 2 U_i (\pi)/\Delta_0$ (b) as
 functions of $b = 2 U_i (\pi)/\Delta$, where $\Delta$ is the order parameter, and $\Delta_0$ is the gap in the absence of impurities.
 The inset in (a) shows the dependence of
T$_c$ on  $b_0$. }
 \label{fig_new}
\end{figure}

The parameter $\Delta$  can be re-expressed in terms of  $\Delta_0
(T)$, which is the BCS gap in the absence of impurities, and  $b_0 =
2 U_i (\pi)/\Delta_0 (T)$, which linearly depends on the impurity
strength. The relation between $\delta = \Delta/\Delta_0$ and $b_0$
(and between $b = b_0/\delta$ and $b_0$) is obtained from the
self-consistent condition on the order parameter
\begin{eqnarray}
&&\Delta = u_{eff} \int_0^{\omega_{max}} \frac{{\bar \Delta}_{\omega_m}}{\sqrt{{\bar \Delta}^2_{\omega_m} + \omega^2_m}},\nonumber \\
&&\Delta_0 = u_{eff} \int_0^{\omega_{max}}
\frac{\Delta_0}{\sqrt{\Delta^2_0 + \omega^2_m}} \label{nn_1}
\end{eqnarray}
where $u_{eff}$ is the normalized interaction in the $s^+$ channel.
Solving these equations, we obtain after some algebra the relations
which express $b_0$ in terms of $b$. They are
\begin{eqnarray}
&& b_0  = b \exp \left( - \frac{\pi b}{4}  \right), ~ b < 1 \nonumber \\
&&b_0 = b  \exp\left( \frac{1}{2}\sqrt{1 - \frac{1}{b^2}} -
\frac{b}{2} \sin^{-1} \frac{1}{b} - \cosh^{-1} b \right), ~ b>1 \nonumber\\
\label{nn_2}
\end{eqnarray}
The first regime corresponds to $\delta = b_0/b  > b_0$, and holds
for $b_0 < e^{-\pi/4} \approx 0.465$. The second regime corresponds
to $b >1$ and describes a gapless superconductivity (${\bar
\Delta}_{\omega_m =0} =0$). When the order parameter $\Delta$ tends
to zero, and $b$ tends to infinity, $b_0$ approaches $0.5$. We plot
$b_0$ vs $b$ in Fig.~\ref{fig_new}(b).

The vanishing of superconductivity at $T=0$ when $b_0$ approaches
$1/2$ also follows from the generic dependence of $T_c$ on the
impurity strength. The calculation parallels the one for an $s-$wave
superconductor with magnetic impurities~\cite{abrgorkov} and yields
\be \ln{\frac{T^0_c}{T_c}} = \psi \left(\frac{1}{2} + \frac{3.53
b_0}{4\pi} \frac{T^0_c}{T_c}\right) - \psi \left(\frac{1}{2} \right)
\label{nn_3} \ee where $\psi (x)$ is the diGamma function. One can
easily check that $T_c$ vanishes when $b_0$ approaches $1/2$. We
plot $T_c (b)$ and $T_c (b_0)$  in Fig. \ref{fig_new}(a).

\subsection{Spin response of an $s^+$ superconductor}

The dynamical spin susceptibility of a superconductor
is given by an RPA-type formula
\begin{equation}
\chi_s ({\bf q}, \Omega) = \frac{\chi^0_s({\bf q},\Omega)}{1 -
\Gamma^{(r)}_{sdw} \chi^0_s ({\bf q}, \Omega)},
\end{equation}
where $\chi^0_s (q, \Omega)$ is the (dimensionless) susceptibility
of an ideal $s^+$ SC (the sum of $GG$ and $FF$ terms with spin
matrices in the vertices). In our case, when $\epsilon^c_{\bf k}
\approx - \epsilon^f_{\bf k+Q}$, and the gap changes sign between
hole and electron FS, one can easily verify that $\chi^0_s ({\bf q}
\approx {\bf Q}, \Omega)$ coincides with the particle-particle
susceptibility for either $c-$ or $f-$ fermions. This leads to
several consequences.

\begin{enumerate}
\item
In the normal state, $\chi^0_s ({\bf Q}, \Omega) = \log{E_F/(-i
\Omega)}$, that is $\mbox{Im} \chi_s ({\bf Q}, \omega)$ only weakly
(logarithmically) depends on frequency. This could be verified in
INS experiments.
\item
In a superconducting state, $\chi_s ({\bf Q}, \Omega)$ has a
resonance below $2\Delta$. Indeed, at $T=0$, in the clean limit and
at small $\Omega$ and ${\bf q-Q}$,
\begin{eqnarray}
&&\chi^0_s ({\bf q}, \Omega) = \log\frac{E_F}{E_0} +
\frac{1}{4\Delta^2} \left(\Omega^2 - v^2 ({\bf q-Q})^2\right)
\label{a_2}
\end{eqnarray}
where $v = v_F/\sqrt{2}$ is the velocity of the Anderson-Bogolyubov
mode in two dimensions (2D), and $E_0$ is the largest of $\Delta$
and the cutoff energy associated with non-equivalence of the two FS.
Substituting this into $\chi_s (q, \Omega)$, and assuming
$\Gamma^{(r)}_{sdw} \log{E_F/E_0} <1$, i.e., no SDW instability, we
find the resonance at $\Omega = \sqrt{(v^2 ({\bf q-Q})^2 +
\Omega^2_0}$, where $\Omega_0 = 2 \Delta (1/\Gamma^{(r)}_{sdw} -
\log E_F/E_0)^{1/2}$. This resonance has been earlier obtained in
the numerical analysis in Refs.~\cite{eremin,scalapino}. It bears
both similarities and differences with the spin resonance in
$d_{x^2-y^2}$ SC. On one hand, both are excitonic resonances, and
both occur because the gap changes sign between the FS points ${\bf
k}$ and ${\bf k+Q}$. On the other hand, the resonance frequency in a
$d_{x^2-y^2}$ SC disperses downwards because of the nodes, while for
a nodeless $s^+$ SC, the resonance disperses upwards, with large
velocity. Indeed, this is only valid if $\Omega \ll 2 \Delta$,
otherwise the dispersion becomes more complex.

Note in passing that, because the two gaps have opposite signs,
there should also exist a resonance mode in the particle-particle
channel, at momentum ${\bf k} = {\bf Q}$, similar to the  Leggett
mode in a two-band superconductor~\cite{Leggett}.

\item
An $s^+$ superconductor has a rather peculiar low-frequency behavior
of $\mbox{Im} \chi_s ({\bf q} \sim {\bf Q}, \Omega \rightarrow 0)$.
In the clean limit,
\begin{equation}
\left. \frac{\mbox{Im} \chi^0_s ({\bf q}, \Omega)}{\Omega}
\right|_{\Omega =0}
 \propto \sum_{\bf k} C_{\bf k,q} \frac{\partial{n_F (E_{\bf k})}}{\partial{E_{\bf k}}},
\label{a_7}
\end{equation}
where $E_{\bf k} = \sqrt{\Delta^2 + \epsilon^2_{\bf k}}$, $n_F(E)$
is Fermi function, and $C_{\bf k, q} = 1 + (\epsilon^c_{\bf k}
\epsilon^f_{\bf k+q} + \Delta^c \Delta^f)/(E^c_{\bf k} E^f_{\bf
k+q})$ is the coherence factor. We see that the coherence factor
vanishes identically for ${\bf q = Q}$ such that $\mbox{Im} \chi^0_s
({\bf q}, \Omega)/\Omega_{|\Omega =0} \propto ({\bf q-Q})^2$.
\end{enumerate}

\subsubsection{NMR spin-lattice relaxation rate}

The spin-lattice relaxation rate measured by nuclear magnetic
resonance (NMR) is given by
\begin{eqnarray}
\frac{1}{T_1} & \propto &  T \sum_{\bf q} \left. \frac{\mbox{Im}
\chi_s ({\bf q}, \Omega)}{\Omega}\right|_{\Omega =0}
\nonumber\\
&\propto & T  \sum_{\bf q} \chi^2_s ({\bf q}, 0)~\left[
\frac{\mbox{Im} \chi^0_s ({\bf q}, \Omega)}{\Omega}\right]_{\Omega
=0}. \label{a_8}
\end{eqnarray}
Because $\chi_s ({\bf q}, \Omega =0)$ is enhanced near ${\bf Q}$,
this region contributes most to the momentum sum. The smallness of
$\mbox{Im} \chi^0_s ({\bf q}, \Omega)/\Omega_{|\Omega =0}$ for ${\bf
q \sim Q}$ then implies that $1/T_1$ has extra smallness in a clean
$s^+$ SC [by the same reason, there is no Hebel-Slichter peak in
$1/T_1$ near $T_c$].

In the presence of impurities, $\mbox{Im} \chi^0_s ({\bf Q},
\Omega)/\Omega_{|\Omega =0}$ remains nonzero, and $1/T_1 \approx T
\mbox{Im} \chi_s ({\bf Q}, \Omega)/\Omega_{|\Omega =0} \int d^2 q
\chi^2_s ({\bf q}, \Omega =0)$. The full expression for $1/T_1$ is
rather involved as one has to include the full $G$ and $F$, {\it
and} the full vertex. It  simplifies considerably if we neglect
vertex corrections and assume that intraband scattering $U_i (0)$
(harmless for superconductivity) well exceeds $\Delta$. In this
case, we obtained analytically, at a finite $T$,
\begin{eqnarray}
\lefteqn{\frac{1}{T_1} = \left.\frac{1}{T_1}\right|_{T_c} \times} &&
\nonumber\\
&& \int_0^\infty \frac{dx}{4 \cosh^2{\frac{x}{2T}}} \left(1 -
\frac{|{\bar \Delta}|^2 -x^2}{\sqrt{(|{\bar \Delta}|^2 -x^2)^2 + 4
x^2 ({\bar \Delta}^{''})^2}}\right) \nonumber \\
\label{a_3}
\end{eqnarray}
where ${\bar \Delta}$ is given by Eq. (\ref{a_4}) with BCS
$T-$dependent $\Delta (T)$. We verified numerically that
 lowest order vertex
corrections do not change the result in any significant way.

\begin{figure}[tbp]
\includegraphics[angle=0,width=0.8\linewidth]{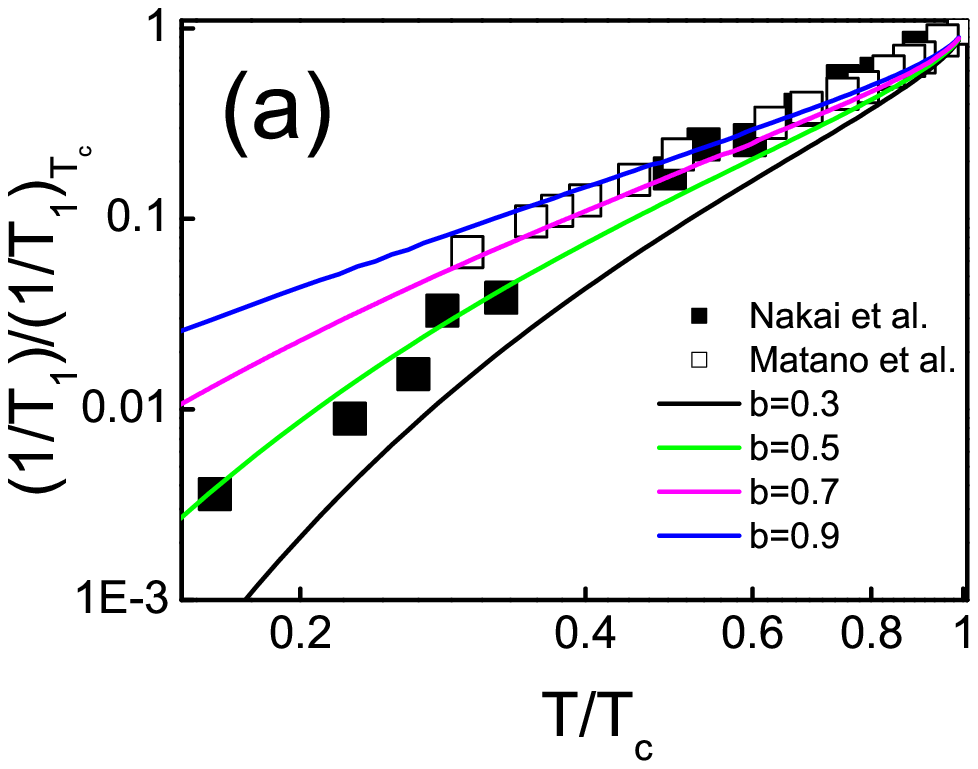}
\includegraphics[angle=0,width=0.8\linewidth]{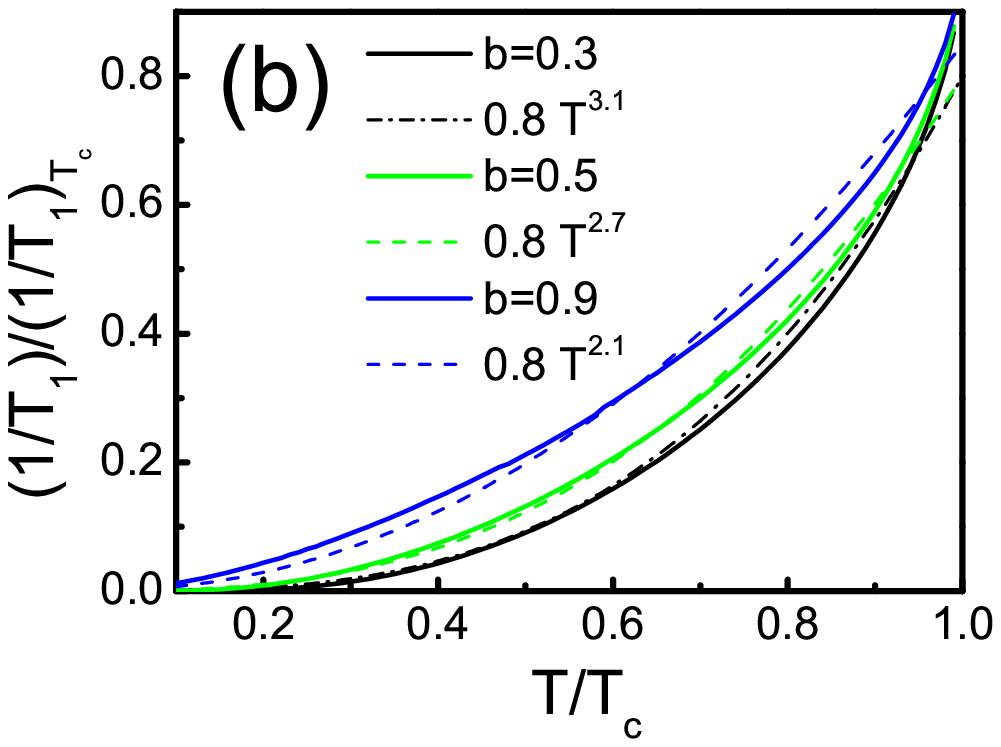}
\caption{(color online) (a) Calculated temperature dependence of
$1/T_1$ for an $s^+$ superconductor with nonmagnetic impurities. The
normalization is chosen such that $1/T_1 =1$ at $T = T_c$. The
theoretical curves are for various values of the parameter $b = 2
U_i (\pi)/\Delta (T=0)$ which measures the strength of the
pair-breaking component of non-magnetic impurities. Gapless
superconductivity occurs for $b >1$. The experimental data are taken
from Refs.\protect\cite{nakai,matano}. (b) Theoretical $1/T_1$ for
different $b$ vs power-law forms . All theoretical dependencies are
exponential in $T$ at very low $T$, but are described by power-laws
$T^\alpha$ over a wide $T$ range below $T_c$. The exponent $\alpha$
decreases as $b$ increases from $\alpha \approx 3$ for $b =0.3$ to
$\alpha \approx 2$ for $b =0.9$. } \label{fig4}
\end{figure}

In Fig.\ref{fig4} we plot the normalized temperature dependence of
$1/T_1 (T)$ for several values of $b = 2 U_i(\pi)/\Delta (T=0)$.
 Stronger impurity scattering corresponds to larger $b$
[it doesn't make a difference
 wthether to parametrize the
impurity strength  in terms of $b$, which depends on impurity strength in a complex way, or $b_0$, which scales linearly with the impurity strength,
because of
one-to-one correspondence between $b$ and $b_0$, see Eq. (\ref{nn_2})].

For
$b<1$, the low-$T$ behavior is exponential, as is expected for a
superconductor without nodes. However, we see that for $b \geq 0.3$,
there is a wide intermediate $T$ range where the behavior of $1/T_1$
closely resembles a power-law $T^\alpha$. The exponent  $\alpha$
decreases as $b$ increases from $\alpha \approx 3$ for $b =0.3$ to
$\alpha \approx 2$ for $b =0.9$.  The $T^3$ behavior was suggested
based on experimental fits and was presented as evidence for
$d-$wave superconductivity in $Fe$-pnictides. Our results show that
$1/T_1 (T)$ in a dirty $s^+$ superconductor mimics a power-law over
a wide $T$ range even when the DOS still vanishes at $\omega =0$,
and $T_c$ is only slightly affected by impurities. Furthermore, we
argue, based on Fig. \ref{fig4}(a) that the experimental $T$
dependence of $1/T_1$ can only approximately be fitted by a
particular power of $T$. We believe that the reported power-law form
reflects intermediate asymptotics of a complex $T$ behavior of
$1/T_1$, and one should reduce temperature further to be able to
distinguish between a true power-law and exponential
behavior~\cite{comm_rev2}.

Note in passing that the theoretical behavior is exponential at the
lowest $T$ only if $b <1$.  For $b=1$, which is the critical $b$ for
a gapless $s^+$ SC, $1/T_1 \propto  T^{5/3}$ at the lowest $T$, and
for larger $b$, $1/T_1 (T) \propto T$.

\begin{figure}[h]
\includegraphics[angle=0,width=0.8\linewidth]{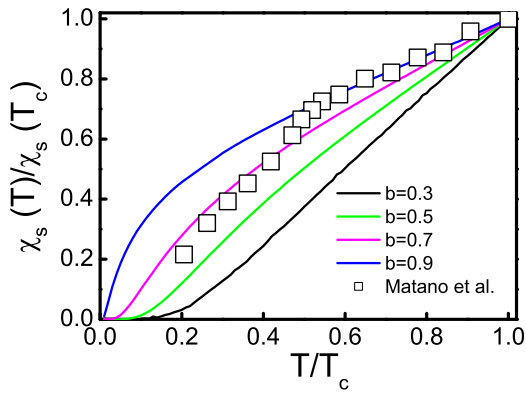}
\caption{(color online) Calculated  temperature dependence
of the uniform susceptibility for various values of $b$.
The experimental data are taken from
Ref.\protect\cite{matano}.} \label{fig5}
\end{figure}

\subsubsection{Uniform susceptibility}

Finally, we also computed uniform spin susceptibility $\chi_s (T)
\approx \chi^0_s (q=\Omega =0)$, measured by Knight shift. It is
obtained by standard means~\cite{gork_rus}, and for a superconductor
with $s^+$ gap symmetry is given by
\begin{widetext}
\begin{eqnarray}
\chi_s (T)  &=&  \chi_s (T_c) \left[1 - \int_0^\infty
\tanh{\frac{x}{2T}}~ \mbox{Im} \frac{{\bar \Delta}^2}{({\bar \Delta}^2
-x^2)^{3/2} - 2U_i (\pi) x^2} \right]\nonumber \\
&=&  \chi_s (T_c) \left[1 - \int_0^\infty
\tanh{\frac{x}{2T}}~ \frac{1}{\Delta (T)}~ \mbox{Im} \frac{1}{(1 - u_x^2)^{3/2} -  b_T}\right]
\label{a_5}
\end{eqnarray}
\end{widetext}
where $u_x = x/{\bar \Delta}$ and ${\bar \Delta}$ depends on $x$ and is given by(\ref{a_4})

We emphasize that ladder series of vertex corrections must be
included in the calculation of $\chi_s (T)$ to recover $SU(2)$ spin
symmetry. Observe that $U_i (0)$ drops from the expression for
$\chi_s (T)$, because impurity scattering of electron within either
hole or electron FS does not differentiate between a conventional
$s-$wave and $s^+$ gap symmetry.

Eq. (\ref{a_5}) is similar, but not identical to the expression for
$\chi_s (T)$ in an ordinary s-wave superconductor with magnetic
impurities~\cite{gork_rus}. In both cases, $\chi_s (T)$ differs from
free-fermion value. However, for magnetic impurities, $\chi_s (T=0)$
becomes finite for any nonzero  strength of the impurity scattering,
while in our case, the impurities are actually non-magnetic, and
$\chi_s (T)$ still vanishes at $T=0$ for all $b <1$, for which the
DOS still vanishes at zero frequency.

We plot Eq. (\ref{a_5}) in Fig. \ref{fig5} for the same $b$ as
$1/T_1$. We see the same trend as in Fig. \ref{fig4}: the
theoretical $T$ dependence of $\chi_s (T)$ is exponential in
$\Delta/T$ at the lowest $T$, but rapidly deviates from exponent
already at small $T$, and is roughly a power-law in $T$ in the same
$T$ range where $1/T_1 (T)$ can be fitted by a power-law. This is
another indication that one should perform Knight shift and $1/T_1$
measurements down very low $T$ to be able to distinguish between the
nodeless $s^+$ state and a SC state with gap nodes. Note also that
the same $b =0.7$ which fits $1/T_1$ data by Matano et
al~\cite{matano} also fits reasonably well their data on the Knight
shift.

\section{Conclusions}

To conclude, in this paper we presented Fermi liquid analysis of SDW
magnetism and superconductivity in $Fe-$pnictides. We considered a
two-band model with small hole and electron pockets located near
$(0,0)$ and ${\bf Q} =(\pi,\pi)$ in the folded BZ. We argued that
for such geometry, particle-hole and particle-particle channels are
nearly identical, and the interactions logarithmically increase at
low energies. We found that the interactions in the SDW and extended
$s-$wave channels ($\Delta_{{\bf k}} =- \Delta_{\bf k+Q}$) become
comparable in strength due to the increase of the intraband pair
hopping term and the reduction of the Hubbard-type intraband
repulsive interaction. We argued that at zero doping, SDW
instability comes first, but at a finite doping, $s^+$
superconducting instability occurs at a higher $T$.

This $s^+$ pairing bears similarity to magnetically mediated
$d_{x^2-y^2}$ pairing in systems with large FS with hot spots in the
sense that in both cases the pairing comes from repulsive
interaction, peaked at ${\bf Q}$, and  requires the gap to change
its sign under ${\bf k} \rightarrow {\bf k+ Q}$. the difference is
that for small pockets, the gap changes sign away from the FS
 and remains constant along the FS.

We analyzed spin response of a clean and dirty  $s^+$ superconductor
and found that (i) it possess a resonance mode which disperses with
the same velocity as Anderson-Bogolyubov mode, (ii) intraband
scattering by non-magnetic impurities is harmless, but interband
scattering affects the system in the same way as magnetic impurities
in an $s-$wave SC, (iii) $1/T_1$ has an extra smallness in the clean
limit due to vanishing of the coherence factor, (iv) in the presence
of impurities, there exists a wide range of $T$ where the
$T-$dependencies of  $1/T_1$ and the uniform susceptibility for an
$s^+$ SC resemble the ones for a SC with nodes.

{\it Note added} While completing this work we became aware that
similar results for spin-lattice relaxation rate, $1/T_1$, in the
superconducting state have been obtained in Ref.
[\onlinecite{parker}].

We acknowledge helpful discussions with Ar. Abanov, J. Betouras, O.
Dolgov, K. Ishida, M. Korshunov, I Mazin, J. Schmalian, Z.
Tesanovic, and G.-q. Zheng. AVC has been supported by NSF-DMR
0604406 and MPI for Physics of Complex Systems in Dresden.


\begin{thebibliography}{99}


\bibitem{kamihara} Y. Kamihara, T. Watanabe, M. Hirano, and H.
Hosono, J. Am. Chem. Soc. {\bf 130} 3296 (2008).
\bibitem{chen1} X.H. Chen, T. Wu,  G. Wu, R.H. Liu, H. Chen, and D.F. Fang, Nature {\bf 453} 761 (2008).
\bibitem{chen2} G.F. Chen, Li Z., Wu D., Li G., Hu W.Z., Dong J., Zheng P., Luo J.L., \and Wang N.L.
    Phys. Rev. Lett. {\bf 100} 247002 (2008).
\bibitem{ren} Z.-A. Ren,
Yang J., Lu W., Yi W., Che G.-C., Dong X.-L., Sun L.-L., \and Zhao Z.-X.
   arXiv:0803.4283 (unpublished).

\bibitem{rotter}  M. Rotter, M. Tegel and D. Johrendt, arXiv:cond-mat/0805.4630 (2008).

\bibitem{dong}   J. Dong, H. J. Zhang, G. Xu, Z. Li, G. Li, W. Z. Hu, D. Wu, G. F. Chen, X. Dai, J. L. Luo, Z. Fang,
N. L. Wang, Europhys. Lett. {\bf 83}, 27006 (2008).

\bibitem{cruz}   Clarina de la Cruz, Q. Huang, J. W. Lynn, J. Li, W. Ratcliff II, J.L. Zarestky,
H.A. Mook, G.F. Chen, J.L. Luo, N.L. Wang, and P. Dai, Nature {\bf
453}, 899 (2008).

\bibitem{nomura} T. Nomura, S.W. Kim, Y. Kamihara, M. Hirano, P.V. Sushko, K. Kato, M. Takata, A.L. Shluger, and
H. Hosono, arXiv:0804.3569 (unpublished).

\bibitem{klauss} H.-H. Klauss, H. Luetkens, R. Klingeler, C. Hess, F.J. Litterst, M. Kraken, M. M. Korshunov,
I. Eremin, S.-L. Drechsler, R. Khasanov, A. Amato, J. Hamann-Borreo,
N. Leps, A. Kondrat, G. Behr, J. Werner, B. B\"uchner, Phys. Rev.
Lett. {\bf 101}, 077005 (2008).


\bibitem{Si} Qimiao Si, and Elihu Abrahams, Phys. Rev. Lett. {\bf 101}, 076401
(2008).

\bibitem{Daghofer} M. Daghofer, A. Moreo, J.A. Riera, E. Arrigoni, D.J. Scalapino, E.
Dagotto, arXiv:0805.0148 (unpublished).

\bibitem{Ma1} F. Ma, and Z.-Y. Lu, Phys. Rev. B {\bf 78}, 033111
(2008).


\bibitem{Raghu}  S. Raghu, X.-L. Qi, C.-X. Liu, D. Scalapino, and S.-C.
Zhang, Phys. Rev. B {\bf 77} 220503(R) (2008).

\bibitem{Hirschfeld1}  C. Cao, P.J. Hirschfeld, H.-P. Cheng, Phys. Rev. B {\bf 77}, 220506(R)
(2008).

\bibitem{Lebegue} S. Leb\`{e}gue, Phys. Rev. B \textbf{75}, 035110 (2007).

\bibitem{Singh} D. Singh and M.-H. Du, Phys. Rev. Lett. {\bf 100}, 237003 (2008).

\bibitem{Boeri} L. Boeri, O.V. Dolgov, and A.A. Golubov, Phys. Rev. Lett. {\bf 101},
026403 (2008).

\bibitem{Mazin} I.I. Mazin, D.J. Singh, M.D. Johannes, and M.H. Du,
Phys. Rev. Lett. {\bf 101}, 057003 (2008).

\bibitem{Kuroki} K. Kuroki,
 S. Onari, R. Arita, H. Usui, Y. Tanaka, H. Kontani, and H. Aoki,
    Phys. Rev. Lett. {\bf 101}, 087004 (2008).

\bibitem{kaminski}  C. Liu, G.D. Samolyuk, Y. Lee, N. Ni, T. Kondo, A.F. Santander-Syro, S.L. Bud'ko,
J.L. McChesney, E. Rotenberg, T. Valla, A. V. Fedorov, P.C.
Canfield, B.N. Harmon, A. Kaminski,  arXiv:0806.3453 (unpublished);
C. Liu, T. Kondo, M.E. Tillman, R. Gordon, G.D. Samolyuk, Y. Lee, C.
Martin, J.L. McChesney, S. Bud'ko, M.A. Tanatar, E. Rotenberg, P.C.
Canfield, R. Prozorov, B.N. Harmon, and A. Kaminski, arXiv:0806.2147
(unpublished);

\bibitem{feng} L.X. Yang, Y. Zhang, H.W. Ou, J.F. Zhao, D.W. Shen, B. Zhou,
J. Wei, F. Chen, M. Xu, C. He, Y. Chen, Z.D. Wang, X.F. Wang, T. Wu,
G. Wu, X.H. Chen, M. Arita, K. Shimada, M. Taniguchi, Z.Y. Lu, T.
Xiang, D.L. Feng, arXiv:0806.2627 (unpublished).

\bibitem{gorkov_barz} V. Barzykin and L.P. Gorkov, JETP Lett. {\bf
88}, 142 (2008).

\bibitem{eremin} M.M. Korshunov, and I. Eremin,  arXiv:0804.1793
(unpublished).

\bibitem{cvetkovic} V. Cvetkovic and Z. Tesanovic, arXiv:0804.4678
(unpublished).

\bibitem{wang} F. Wang, H. Zhai, Y. Ran, A. Vishwanath, and D.-H.
Lee, arXiv:0805.3343 (unpublished)

\bibitem{wang_2} F. Wang, H. Zhai, Y. Ran, A. Vishwanath, and D.-H.
Lee,  arXiv:0807.0498 (unpublished)

\bibitem{lorenziana} J. Lorenzana, G. Seibold, C. Ortix, and M.
Grilli, arXiv:0807.2412v1 (unpublished).

\bibitem{kondo} T. Kondo, A.F. Santander-Syro, O. Copie, C. Liu, M.E. Tillman, E.D. Mun,
J. Schmalian, S.L. Bud'ko, M.A. Tanatar, P.C. Canfield, and A.
Kaminski, arXiv:0807.0815 (unpublished).

\bibitem{ding} H. Ding, P. Richard, K. Nakayama, T. Sugawara, T. Arakane, Y. Sekiba, A. Takayama,
S. Souma, T. Sato, T. Takahashi, Z. Wang, X. Dai, Z. Fang, G.F.
Chen, J.L. Luo, and N.L. Wang , Europhys. Lett. {\bf 83}, 47001
(2008).

\bibitem{kaminski_gap} T. Kondo, A.F. Santander-Syro, O. Copie, C. Liu, M.E. Tillman, E.D. Mun,
J. Schmalian, S.L. Bud'ko, M.A. Tanatar, P.C. Canfield, and A.
Kaminski,  arXiv:0807.0815 (unpublished).

\bibitem{chen} T. Y. Chen, Z. Tesanovic, R. H. Liu, X. H. Chen, C. L.
Chien, Nature {\bf 453}, 1224 (2008).

\bibitem{nakai}  Y. Nakai, K. Ishida, Y. Kamihara, M. Hirano, and H. Hosono
J. Phys. Soc. Jpn. {\bf 77}, 073701 (2008).

\bibitem{grafe}  H.-J. Grafe, D. Paar, G. Lang, N.J. Curro, G. Behr, J. Werner, J. Hamann-Borrero,
C. Hess, N. Leps, R. Klingeler, and B. Buechner,  Phys. Rev. Lett.
{\bf 101}, 047003 (2008).

\bibitem{matano}  K. Matano, Z.A. Ren, X.L. Dong, L.L. Sun, Z.X. Zhao, and G.-q.
Zheng, Europhys. Lett. {\bf 83}, 57001 (2008).

\bibitem{subir_1} Cenke Xu, Markus Mueller, and Subir Sachdev, Phys. Rev. B 78, 020501(R) (2008);
Cenke Xu, Yang Qi, and Subir Sachdev, arXiv:0807.1542.

\bibitem{com_rev}
Note that our two-band model differs conceptually
from the two-band model based on the $d_{yz}$ and $d_{xz}$ orbitals considered in \protect\cite{Raghu}. Band structure
calculations\cite{Boeri} show that not only $d_{yz}$ and $d_{xz}$ orbitals, but  all Fe 5$d$-orbitals contribute to
the bands crossing the Fermi level, and, besides,
 Fe 5$d$-orbitals  strongly hybridize with the As $p$-states.
The tight-binding model of the pnictides then should
 include  8 bands in the unfolded BZ.
Our model with two hybridized orbitals and two bands crossing the Fermi level
 is not related to actual  Fe 5$d$ and  As $p$ -orbitals, and
 is simply a phenomenological minimal model to describe SDW and superconductivity.

\bibitem{phillips} J. Wu, P. Phillips, and A.H. Castro-Neto, arXiv:0805.2167.

\bibitem{landau}  A.\ A.\ Abrikosov, L.\ P.\ Gorkov, and
I.\ E.\ Dzyaloshinski, \emph{Methods of quantum field theory in
statistical physics}, (Dover Publications, New York, 1963); E. M.
Lifshitz
and L. P. Pitaevski, \emph{%
Statistical Physics}, (Pergamon Press, 1980).

\bibitem{dzyal} A.T. Zheleznyak, V.M. Yakovenko, and I.E.
Dzyaloshinskii, Phys. Rev. B {\bf 55}, 3200 (1997).

\bibitem{mcmillan}W.L. McMillan, Phys. Rev. {\bf 167}, 331 (1968).

\bibitem{quantum} E. Shender, Sov. Phys. JETP, {\bf 56}, 178 (1982);
A. Pimpinelli, E. Rastelli, and A. Tassi, J. Phys.: Cond. Matt {\bf
1}, 2131 (1989); A. Moreo {\it et al.}, \prb {\bf 42}, 6283 (1990);
A. Chubukov, \prb {\bf 44}, 392 (1991).

\bibitem{coleman} P. Chandra, P. Coleman, and A.I. Larkin, \prl {\bf 64}, 88 (1990).

\bibitem{hirschfeld11} For recent literature see e.g., T. Dahm, P. J. Hirschfeld,
D. J. Scalapino, and L. Zhu, Phys. Rev. B {\bf 72}, 214512 (2005).


\bibitem{golubov} A.A. Golubov, and I.I. Mazin,  Phys. Rev. B {\bf 55}, 3200
(1997).

\bibitem{preosti}  G. Preosti, and P Muzikar, Phys. Rev. B
{\bf 54}, 3489 (1996).

\bibitem{abrgorkov} A.A. Abrikosov and L.P. Gorkov, Sov. Phys. JETP {\bf 12}, 1243
(1961).


\bibitem{scalapino}  T.A. Maier, and D.J. Scalapino, Phys. Rev. B {\bf 78}, 020514(R) (2008).

\bibitem{Leggett} A.J. Leggett, Prog. Theor. Phys. {\bf 36}, 901
(1966).

\bibitem{comm_rev2}

The behavior of $1/T_1$ is further complicated by the presence of
four Fermi surfaces and the variation of the gap size between the
two Fermi surfaces near the same momenta, as evidenced by ARPES
measurements\protect\cite{ding}.

\bibitem{gork_rus} L.P. Gorkov and A.I. Rusinov, Sov. Phys. JETP,
{\bf 19}, 922 (1964); W. A. Roshen and J. Ruvalds, Phys. Rev. B {\bf
31}, 2929 (1985).

\bibitem{parker} D. Parker, O. Dolgov, M.M. Korshunov, A. Golubov,
and I.I. Mazin, arXiv:0807.3729 (unpublished).


\end{thebibliography}
\end{document}